\documentclass[aps,pre,preprint,showpacs]{revtex4}

\usepackage{graphicx}
\usepackage{latexsym} 
\usepackage{amsmath,amscd}
\usepackage{times}

\begin{document}

\title{\bf A growth walk model for estimating the canonical partition function of Interacting Self Avoiding Walk }
\author{S. L. Narasimhan$^{*}$ and P. S. R. Krishna}
\affiliation{
Solid State Physics Division, Bhabha Atomic Research Centre, Mumbai - 400085, India.}
\author{M. Ponmurugan$^1$ and K. P. N. Murthy$^{1,2}$}
\affiliation{
$^1$ Materials Science Division, Indira Gandhi Center of Atomic Reasearch, Kalpakkam - 603102, Tamil Nadu, India.\\
$^2$ School of Physics, University of Hyderabad, Central University P.O, Gachibowli, Hyderabad - 500046,   Andhra Pradesh, India.}

\date{\today}
 

\begin{abstract}

We have explained in detail why the canonical partition function of Interacting Self Avoiding Walk (ISAW), is exactly equivalent to the configurational average of the weights associated with growth walks, such as the Interacting Growth Walk (IGW), if the average is taken over the entire genealogical tree of the walk. In this context, we have shown that it is not always possible to factor the the density of states out of the canonical partition function if the local growth rule is temperature-dependent. We have presented Monte Carlo results for IGWs on a diamond lattice in order to demonstrate that the actual set of IGW configurations available for study is temperature-dependent even though the weighted averages lead to the expected thermodynamic behavior of Interacting Self Avoiding Walk (ISAW).

\end{abstract}

\maketitle

\section{Introduction}

Monte Carlo simulations, based on Metropolis sampling algorithm, have immensely contributed to our understanding of a variety of complex physical systems and their thermodynamic behavior ~\cite{DP}. Yet, they are also  known to be computationally inefficient in situations where low entropy microstates of a canonical system are to be sampled in sufficient numbers in order to obtain accurate estimates of relevant thermodynamic parameters. Employing suitable biassing rules that favor these microstates, especially in irreversible growth models, does not ensure a neat solution to this problem because the associated weights could wildly fluctuate leading to large errors. Wang-Landau flat histogram algorithm ~\cite{WL} is a recently proposed dynamical method for efficiently sampling such microstates and hence, for accurately computing the Density of States (DoS) of the system under study.

A particularly interesting idea, highlighted recently by Prellberg and Krawczyk ~\cite{PK} in the context of Self Avoiding Walk (SAW) ~\cite{deG}, is to recognize the DoS as simply the average of weights associated with all Monte Carlo attempts to sample the required microstates; acceptance of a microstate depends on how the associated weight compares with the average and is implemented in such a way that the energy histogram becomes progressively flatter. In this step-by-step growth model, also known as the Kinetic Growth Walk (KGW) ~\cite{stan}, the weight associated with a microstate (a walk configuration) is the product of single step weights which are, in fact, the local {\it microcanonical} partition functions or equivalently the number of available directions for the individual steps. Since the model is athermal, it is intuitively clear that the average weight could provide an estimate of the DoS.

If, on the other hand, the individual steps are sampled on the basis of the {\it energy} being gained by the walk, their associated weights are customarily taken to be equal to the inverse of the corresponding step probabilities; again the configurational average of their products could lead to an estimate of the DoS. 
An interesting alternative is to set the single step weights equal to the local {\it canonical} partition functions, a straightforward generalization of the athermal case. Assuming that the configurational average of their products over all Monte Carlo attempts leads to an estimate of the {\it canonical} partition function, it is not necessary that it would be a sum of terms that are factorizable into the (athermal) DoS and the corresponding Boltzmann factor. 

In this paper, we clarify this point by using Interacting Growth Walk (IGW) model ~\cite{SLN}, which is a finite temperature generalization of the KGW. We explain how the {\it canonical} partition function for walks of given length is exactly equal to the average of the products of local partition functions associated with all possible walk configurations including those that have failed to grow to the full length. In this case, we show that the density of states (DoS) cannot be factored out of the canonical partition function. Further, we present Monte Carlo results for IGWs on the diamond lattice and show that the set of IGW configurations available for study is temperature-dependent, in contrast to the athermal set of KGW configurations, even though the weighted averages lead to the expected thermodynamic behavior of SAW.

\section{Step-by-step growth of a self-avoiding walk}

Consider a SAW configuration, ${\cal C}_{K-1}$, made up of an ordered set of $K-1$ directions, $\{ \mu _1,\mu _2,\cdots ,\mu _{K-1}\}$, taken consecutively on a regular lattice of coordination number $z$. Recognizing the fact that self-avoidance for the current step is a non-local requirement involving the entire walk configuration, we denote by $a_K({\cal C}_{K-1})$ the number of acceptable or available directions for the $K^{th}$ step. The walk proceeds further only if the direction chosen for the next step is acceptable; it is 'trapped' at the $(K-1)^{th}$ step and will not grow further, if $a_K({\cal C}_{K-1}) = 0$. In other words, local acceptability criteria decide the length and configuration of the walk.

It is quite likely that an acceptable step leads to a site some of whose nearest neighbours are sites through which the walk configuration has already grown. Let $n(\mu _K; {\cal C}_{K-1})$ be the number of such non-bonded nearest neighbours, also called {\it contacts}, encountered by the $K^{th}$ step in the direction $\mu _K$. Clearly, the total number of {\it contacts} in the configuration ${\cal C}_K$ is the sum of {\it contacts} made by each step in the walk. By assigning a quantum of energy, say $\epsilon$, to each of these {\it contact}s, we will be able to treat the walk as a thermal object.  

There are $z$ possible directions for the first step taken from an arbitrary lattice site, called the 'origin'; each of these directions leads to $(z-1)$ possible directions for the second step, and so on until the first $M_z$ steps of the walk are taken without making any contact. The total number of possible configurations identified upto $M_z$ steps is therefore $B_{M_z} \equiv z(z-1)^{M_z-1}$.

Some of these configurations would make {\it contacts}, for the first time, at the $(M_z+1)^{th}$ step. For example, $M_z = 4, 2$ and $1$ on honecomb, square and triangular lattices respectively. Subsequently, each of these contact making configurations would have, say, $a_{M_z+2}({\cal C}_{M_z+1})$ directions for the $(M_z+2)^{th}$ step. Some of these growing configurations will be geometrically 'trapped' at the $N_z^{th}$ step and hence cannot grow further. The minimum value of $N_z$ at which trapping occurs for the first time depends on the lattice on which the walk is grown. For example, min${N_z} = 9, 7$ and $6$ on honeycomb, square and triangular lattices respectively.  

If we map out all the configurations till they are either geometrically trapped or have grown to their full specified length, say $N$, we have the {\it genealogical tree}, ${\cal Z}_N$, of SAWs of length less than or equal to $N$. Clearly, any trapped $K$-step SAW configuration ($K<N$), realizable by a growth algorithm, is a {\it branch} of the {\it genealogical tree}. It flowcharts all possible outcomes of a growth algorithm and is, in fact assumed to be made up of all possible $N$-step SAWs as well as all possible trapped $K$-step SAWs ($K<N$). The shortest branch of the tree is of length $N_z$, at which length the growth of the branch is stopped on a given lattice. Counting the number of branches of length $K$ in the tree, ${\cal Z}_N$, for all $K$ in range $[N_z,N]$ could provide a  complete description of the treeIt is of interest to know whether a growth algorithm that uses local growth rules to trace out a branch can estimate, in particular, the number of branches of length $N$, which is equal to the total number of $N$-step SAWs, $Z_N$.

Let $P({\cal C}_K,\beta )$ be the probability of growing a $K$-step walk, or equivalently of realizing a $K$-step branch of the genealogical tree ${\cal Z}_N$, at an inverse temperature $\beta$. Since any branch of this tree is a either a realizable SAW configuration of length equal to $N$ or a realizable trapped SAW configuration of length less than $N$ , and {\it vice versa}, we have the normalization for their growth probabilities:
\begin{equation}
\sum _{K=N_z}^N \left( \sum _{{\cal C}_K} P({\cal C}_K,\beta )\right) = 1
\end{equation}
Since ${\cal Z}_N$ ($N > N_z$) does not have a branch of length less than $N_z$, $P({\cal C}_{K<N_z},\beta ) = 0$ and hence the summation is from $N_z$ onwards. 

In terms of the single step probabilities, $p(\mu _L ; {\cal C}_{L-1},\beta)$, we can write
\begin{equation}
P({\cal C}_K,\beta ) = \left\{ 
                      \begin{array}{cc}
                       B_{M_z}^{-1}\prod _{L=M_z+1}^K p(\mu _L ; {\cal C}_{L-1},\beta) &   
                                \mbox{ if all the steps are acceptable}\\
                       0 & \mbox{  if } L^{th}\mbox{ step is not acceptable}(L<K)
                      \end{array}
                     \right.      
\end{equation} 
where the prefactor $B_{M_z}^{-1} \equiv [z(z-1)^{M_z-1}]^{-1}$ is the probability of growing the initial segment of length $M_z < N_z$, which does not make any contact during its growth and hence is athermal.

Single step probabilities, $p(\mu _L ; {\cal C}_{L-1},\beta)$, are in general assumed to be temperature dependent so as to take into account the possibility of bias due to contacts. However, in the case of KGW, it is temperature-independent and is given by
\begin{equation}
p(\mu _L ; {\cal C}_{L-1}) = \frac{1}{a_L({\cal C}_{L-1})};\quad a_L({\cal C}_{L-1}) > 0
\end{equation}
It is locally normalized over all acceptable steps. In the case of IGW, on the other hand, $p(\mu _L ; {\cal C}_{L-1},\beta)$ is a temperature-dependent locally normalized jump probability for the $L^{th}$ step in the direction $\mu _L$:
\begin{equation}
p(\mu _L ; {\cal C}_{L-1},\beta)  \equiv  \frac{e^{\beta n(\mu _L; {\cal C}_{L-1})\epsilon}}
                                              {\sum _{\mu _L = 1}^{a_L({\cal C}_{L-1})}
                                               e^{\beta n(\mu _L; {\cal C}_{L-1})\epsilon}}
                                               ;\quad a_L({\cal C}_{L-1}) > 0                                               
\end{equation}
where $a_L({\cal C}_{L-1})$ is the number of available sites for the $L^{th}$ step and, without loss of generality, $\epsilon$ may be set equal to unity.

\section{Estimating the canonical partition function using the growth probabilities}

Growing a SAW of length more than the minimum $M_z$ steps with history-dependent, local step-probabilities has an important consequence - namely, that its growth probability, $P$, will not be the same as that obtained by growing the same configuration in the reverse order. The reason for this is the different local growth environments in which the walk has to sample its next step. One way of quantifying and correcting for this is to assign a {\it canonical} weight to the $L^{th}$ step in the direction $\mu _L$ of a growing configuration ${\cal C}_{L-1}$, if acceptable ({\it i.e.,} if $a(\mu _L ;{\cal C}_{L-1}) = 1$),
\begin{equation}
\omega (\mu _L;{\cal C}_{L-1},\beta) \equiv \frac{e^{\beta n(\mu _L;{\cal C}_{L-1})}}
                                           {p(\mu _L;{\cal C}_{L-1},\beta)};
                                           \quad 1 < L \leq N
\end{equation}
It may be recognized immediately as Grassberger's PERM-B weight ~\cite{Grass} for IGW and is, of course, zero for a step that is not acceptable. These weights assigned to the individual steps, in turn, define a {\it canonical} weight that can be assigned to an $N$-step configuration, ${\cal C}_{N}$:
\begin{eqnarray}
W _N({\cal C}_N,\beta) & = & B_{M_z}\prod _{L=M_z+1}^{N}\omega (\mu _L;{\cal C}_{L-1},\beta) \\
                    & = & \frac{e^{\beta n({\cal C}_N)}}{P({\cal C}_N,\beta )}\quad \quad 
                                                                      \mbox{  (by Eq.(2))}  
\end{eqnarray}
where the prefactor $B_{M_z} \equiv z(z-1)^{M_z-1}$ is due to the fact that the walk does not make any contacts for the first $M_z$ steps, and $n({\cal C}_N)$ ( $\equiv \sum _{L=M_z+1}^{N} n(\mu _L;{\cal C}_{L-1})$ ) is the total number of contacts in an $N$-step configuration ${\cal C}_N$. Note that $W _N({\cal C}_N,\beta) = 0$, by definition, for walks of lengths less than $N$.  

The average value of $W _N({\cal C}_N,\beta)$, taken over the entire genealogical tree ${\cal Z}_N$, is given by
\begin{eqnarray}
\langle W _N(\beta)\rangle & \equiv & 
         \frac{\sum _{L=1}^N \left( \sum _{{\cal C}_L}W _N({\cal C}_N,\beta)P({\cal C}_L,\beta)                   
                             \right)}
              {\sum _{L=1}^N \left( \sum _{{\cal C}_L}P({\cal C}_L,\beta)\right)}\\
                          & = & 
               \sum _{{\cal C}_N}W _N({\cal C}_N,\beta)P({\cal C}_N,\beta)
\end{eqnarray}
The last identity is due to the normalization of the growth probabilities given by Eq.(1), as well as due to $W _N({\cal C}_N,\beta) = 0$, by definition, for walks of length less than $N$. From Eq.(7), it is clear that
\begin{equation}
\langle W _N(\beta)\rangle = \sum _{{\cal C}_N}e^{\beta n({\cal C}_N)} \equiv Z_N(\beta)
\end{equation}
where $Z_N(\beta)$ is the {\it canonical Partition Function} for $N$-step Interacting Self Avoiding Walks (ISAW) ~\cite{HS}. In other words, averaging the canonical weights, defined in Eq.(7), over the entire {\it genealogical tree}, ${\cal Z}_N$, gives the exact value of $Z_N(\beta)$. 

Because mapping the entire genealogical tree, ${\cal Z}_N$, is quite a formidable task for large values of  $N$, Monte Carlo methods are employed for estimating the average value of $W _N(\beta)$:
\begin{equation}
\overline{W _N(\beta)}_S = \frac{\sum _{{\cal C}_N} W _N({\cal C}_N,\beta)}{S}
\end{equation}
where the summation is over all successful $N$-step walks and $S$ is the total number of attempts made to generate them. With the probability of growing a configuration implicitly taken care by the algorithmic rules, this equation may be recognized as an equivalent of Eq.(8); the Monte Carlo estimate of $W _N(\beta)$ will approach that given by Eq.(10) as the number of attempts, $S$, becomes sufficiently large ({\it i.e.,} $\overline{W _N(\beta)}_{S\to \infty} \to \langle W _N(\beta)\rangle $). 

It must be noted that the identity, Eq.(10), holds good whatever be the growth walk used. For example, the single-step probability of KGW is temperature-independent (Eq.(3)) whereas that of IGW is temperature-dependent (Eq.(4)), yet either of them could be used for estimating $Z_N(\beta)$. In order to understand the basic difference between them, we first rewrite Eq.(9) in the form,
\begin{eqnarray}
\langle W _N(\beta)\rangle & = & \sum _{n=0}^{n_X(N)} \left(
                                  \sum _{{\cal C}_{N,n}}
                                        W_N ({\cal C}_{N,n},\beta)P({\cal C}_{N,n},\beta)\right)\\
                                & \equiv & \sum _{n=0}^{n_X(N)}
                                        \langle W _{N,n}(\beta)\rangle 
\end{eqnarray}
where the inner summation is over all the $N$-step configurations that make $n$ {\it contacts} and $n_X(N)$ is the maximum number of {\it contacts} an $N$-step configuration will make. A comparison with Eq.(10) immediately leads to the identification,
\begin{equation}
\sum _{n=0}^{n_X(N)}\langle W _{N,n}(\beta)\rangle = Z_N(\beta) \equiv 
                                                     \sum _{n=0}^{n_X(N)}g_N(n)e^{\beta n}
\end{equation}
where $g_N(n)$ denotes the total number of configurations making $n$ {\it contacts}, or equivalently the Density of States (DoS). This begs the question whether a term-by-term identification is also implied. 

In the case of KGW, the weights being temperature-independent, we have the identity,
\begin{equation}
\langle W_{N,n}\rangle = \sum _{{\cal C}_{N,n}}W_N ({\cal C}_{N,n})P({\cal C}_{N,n})\equiv g_N(n)
\end{equation}
that leads to the factorizability of the individual terms $\langle W _{N,n}(\beta)\rangle $, namely,
\begin{equation}
\langle W _{N,n}(\beta)\rangle \equiv \langle W_{N,n}\rangle e^{\beta n}
\end{equation}
In other words, term-by-term identification in Eq.(14) is meaningful for KGW because the athermally generated configurations are assigned appropriate Boltzmann factors {\it a posteriori}.

On the contrary, the probability of generating an IGW configuration is temperature-dependent, and from Eqns. (4 - 7), we have the product,
\begin{equation}
W_N ({\cal C}_{N,n},\beta)P({\cal C}_{N,n},\beta )  =  \left\{ B_{M_z}\prod _{L=M_z+1}^{N}\left( 
                                         \sum _{\mu _L = 1}^{a_L({\cal C}_{L-1})}
                                          e^{\beta n(\mu _L; {\cal C}_{L-1})\epsilon} \right) \right\}   
                              \delta \left( n - n({\cal C}_N)\right)
\end{equation}
where $\delta $-function ensures that the configuration ${\cal C}_N$ has $n$ {\it contacts}. It is clearly not factorizable into temperature-independent (DoS) and temperature-dependent terms, like in the case of KGW (Eqs.(15 and 16)). Yet, Eqs.(10 and 12) ensure that the configurational average of IGW weights (or equivalently, the PERM-B weights) provides an estimate of the canonical partition function and hence can also be used for computing the energy fluctuation as a function of temperature. 

A Monte Carlo estimate of $\langle W _{N,n}(\beta)\rangle $ follows from Eq.(11):
\begin{equation}
\overline{W _{N,n}(\beta)}_S = \frac{\sum _{{\cal C}_{N,n}} W _N({\cal C}_{N,n},\beta)}{S}
\end{equation}
In the case of KGW, the weights are temperature-independent and we immediately have a Monte Carlo estimate of the Density of States, $g_N(n)$, from Eq.(15):
\begin{equation}
\overline{g_N(n)}_S = \frac{\sum _{{\cal C}_{N,n}}W_N({\cal C}_{N,n})}{S} 
\end{equation}
In the case of IGW, however, such a direct estimate of the DoS is not possible.  

A dynamical algorithm for estimating $\overline{W _K(\beta)}_S$ at any given step $K = 1, 2,\cdots , N$ has recently been proposed by Prellberg and Krawczyk ~\cite{PK}. Let $W _K({\cal C}_K,\beta)$ be the canonical weight assigned to the $(S+1)^{th}$ configuration, ${\cal C}_K$, at the $K^{th}$ step. This configuration is either allowed to grow further or is terminated depending on how its weight compares with the updated average, $\overline{W _K(\beta)}_{S+1} = [S\overline{W _K(\beta)}_S + W _K({\cal C}_K,\beta)]/(S+1)$:
\begin{equation}
r_K({\cal C}_K,\beta) \equiv \frac{W _K({\cal C}_K,\beta)}
                                  {\overline{W _K(\beta)}_{S+1}} = \left\{
                                      \begin{array}{cc}
                                            < 1 & \quad \mbox{ continue the walk with probability r}\\
                                            \geq 1 & \quad \mbox{ continue (enrich) the walk}
                                      \end{array}
                                                                     \right.
\end{equation}
Since $\overline{W _K(\beta)}_S$ is an estimate of the canonical partition function, $Z_K$, the parameter $r_K({\cal C}_K,\beta)$ could be considered as the Boltzmann factor $e^{\beta \Delta F({\cal C}_K,\beta)}$, where $\Delta F({\cal C}_K,\beta) = \beta ^{-1}\log [W _K({\cal C}_K,\beta)/ \ \overline{W _K(\beta)}_{S+1}]$. Hence, {\it without enrichment}, Eq.(20) is a simple Metropolis criterion that is also physically meaningful for the growth walks.


\section{A Monte Carlo example}

Using this simple Metropolis criterion, without enrichment, we have estimated, and shown in Fig.1, normalized fluctuations in the number of contacts per monomer, $\sigma ^2(m)/N$, for fairly short IGWs of length upto $256$ steps on a diamond lattice. As the walk length increases, the peak shifts towards the expected value, $\beta \sim 0.44$ ~\cite{Baum}, indicating thereby that the specific heat data are not sensitive to the factorizability of the canonical weights associated with KGW and IGW respectively.

%
%
\begin{figure}
\includegraphics[width=4.0in,height=3.5in]{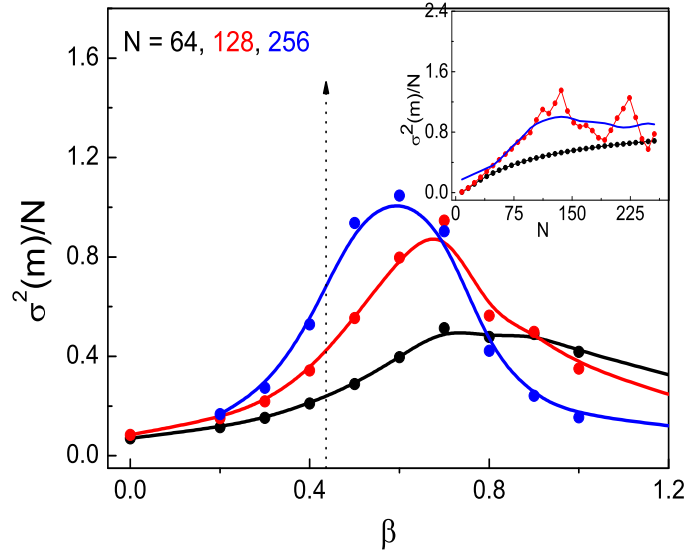}
\caption{Normalized weighted variance, $\sigma ^2(m)/N$, per monomer as a function of the inverse temperature, $\beta$ for IGWs on the diamond lattice for lengths $N = 64, 128$ and $256$. The peak shifts towards the expected value $\beta _{\theta} \sim 0.44$ for longer walks. INSET: Normalized variance per monomer plotted as a function of the length of the walk grown at $\beta = 0.4$. The unweighted data is monotonically increasing with $N$, whereas the weighted ones are highly fluctuating. A simple $10$-point adjacent averaging of the unweighted data is shown by the continuous line. }
\end{figure}

Large fluctuations in the weights could be a severe problem, as illustrated in the inset of Fig.1 for example. In fact, sample loss due to pruning outweighs the reduction in loss due to attrition at higher values of $\beta$; consequently, reduction in the effective sample size may also contribute to wild fluctuation in the weights. The data presented in the inset correspond to IGWs grown at $\beta = 0.7$; it took about three hours of Pentium IV $3.2$ GHz processor time to collect them. While the unweighted variance per monomer increases monotonically with the length of the walk, the weighted ones fluctuate wildly; nevertheless, a definite trend could still be discerned by a simple adjacent averaging of the weighted data. 

%
%
\begin{figure}
\includegraphics[width=4.0in,height=3.5in]{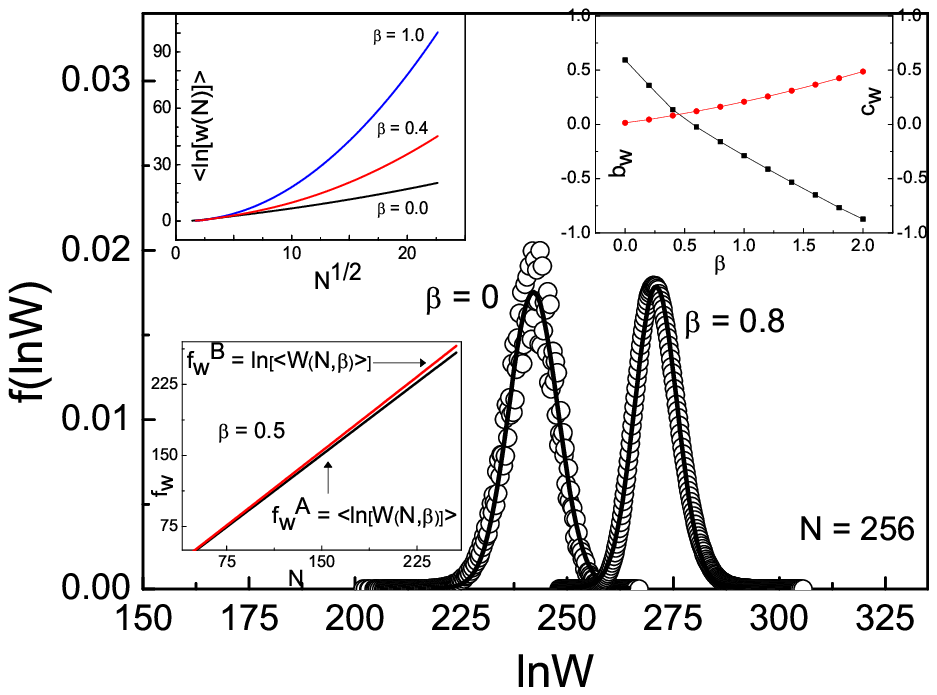}
\caption{Normalized distribution of logarithm values of the IGW (or PERM-B) weights for $N = 256$ at $\beta = 0$ and $0.8$. UPPER LEFT INSET: Peak values of the distribution,$\langle ln[W(N,\beta)]\rangle$, as a function of ${\sqrt N}$ for $\beta = 0, 0.4$ and $1.0$ (bottom to top). The data fall on a parabola, $\langle ln[W(N,\beta)]\rangle \sim b_w {\sqrt N} + c_w N$. RIGHT INSET: The parameters, $b_w$ and $c_w$, versus $\beta$. LOWER LEFT INSET: A demonstration of the fact that $\langle ln[W(N,\beta)]\rangle \neq ln[\langle W(N,\beta)\rangle]$ at $\beta = 0.5$. The deviation becomes larger at higher values of $N$.  }
\end{figure}

The overflow problems plaguing the IGW weights for long walks are taken care of by normalizing the canonical partition function for the current step with the effective coordination number at a given value of $\beta$, which is nothing but the asymptotic estimate of the configuration-averaged, local canonical partition function per step. It is observed that the weights are log-normally distributed (Fig.2), and its peak value, $\langle ln[W(N,\beta)]\rangle$, is asymptotically proportional to $N$ for any given value of $\beta$. In fact, we find that $\langle ln[W(N,\beta)]\rangle = a_w(\beta) + b_w(\beta){\sqrt N} +c_w(\beta)N$ (upper left inset of Fig.2), where the small scale-shift parameter, $a_w(\beta)$, may be ignored for large $N$; $\beta$-dependence of the parameters $b_w$ and $c_w$ is presented in the right inset of Fig.2.
So, in the asymptotic limit ($N\to \infty$), $\langle ln[W(N,\beta)]\rangle \propto N$ which is consistent with the expected scaling form $\langle W(N,\beta)\rangle \sim z_{eff}^N$, where $z_{eff}$ is the effective coordination number of the lattice. It must, of course, be mentioned here that $\langle ln[W(N,\beta)]\rangle \neq ln[\langle W(N,\beta)\rangle]$, even though they both vary linearly with $N$ for long walks, as demonstrated in the lower left inset of Fig.2 for $\beta = 0.5$.

%
%
\begin{figure}
\includegraphics[width=4.0in,height=3.5in]{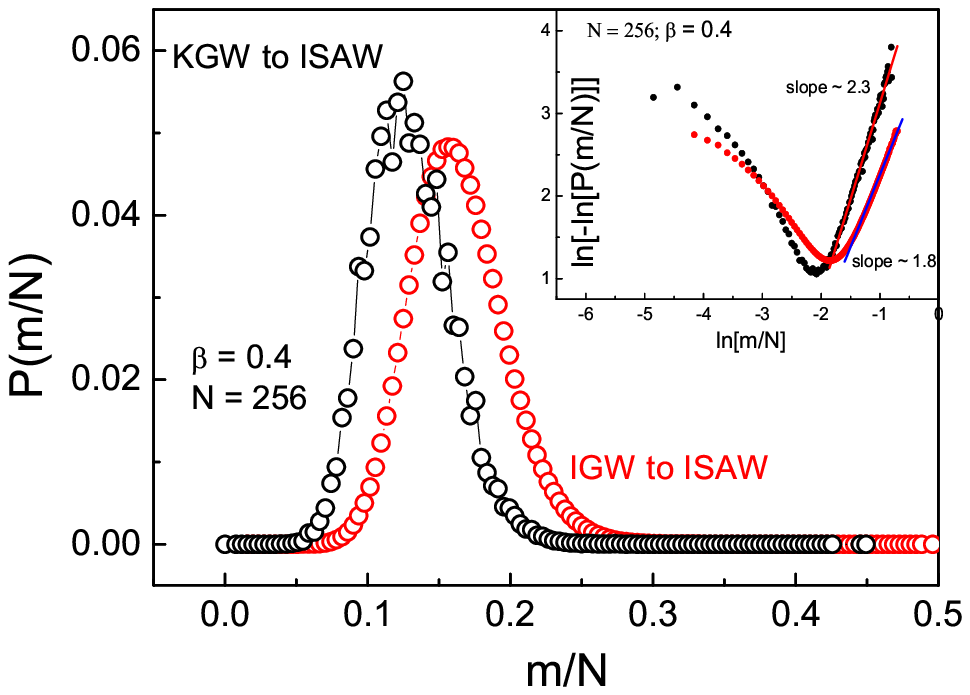}
\caption{Normalized probability distribution of the fraction of contacts for walks of length $N = 256$ at $\beta = 0.4$. The one peaking at a smaller value of $m/N$ corresponds to KGW canonically reweighted to $\beta = 0.4$. The other one peaking at a higher value of $m/N$ corresponds to IGW grown at $\beta = 0.4$ and (PERM-B) weighted. INSET: Same data, but $ln[-ln[P(m/N)]]$ is shown as a function of $ln[m/N]$.  }
\end{figure}

On the other hand, the weighted distribution for the number of contacts per monomer, $P(m/N)$, is quite sensitive to the the type of growth walk used. This is evident from Fig.3 in which the weighted $P(m/N)$ for IGW grown at $\beta = 0.4$ is compared with the distribution for KGW canonically reweighted at $\beta = 0.4$. It is clear that the IGW configurations are more compact than the KGW configurations, eventhough they both have been counted as ISAW configurations at $\beta = 0.4$. The distribution has a non-Gaussian functional form, $P(m/N) \sim -(m/N)^{\alpha}$, especially away from the peak region, as is evident from the inset of Fig.3. This is in agreement with Baumgaertner's analysis ~\cite{Baum}.

\section{Summary}

We have explained in detail how the configurational average of the weights associated with growth walks, such as IGW, provides an esimate of the canonical partition function. We have drawn attention to the fact that the individual terms of the canonical partition function are not expressible as products of the density of states and the Boltzmann factor if the local growth rule is temperature-dependent. In fact, the  parameter, $\beta$, that tunes the step-by-step growth of an IGW may be interpreted {\it a posteriori} as the inverse 'bath' temperature if the growing configuration is assigned a PERM-B weight, as given by Eq.(5). Numerical support for this is provided by the specific heat data. Yet, the actual set of configurations generated depends very much on whether the growth rule employed is athermal or not. For example, at large values of $\beta$, the IGW growth rule may not generate certain (compact) configurations, which are expected to be realizable if $\beta ^{-1}$ were a bath temperature. We have demonstrated some of these subtle points by presenting Monte Carlo results obtained for IGWs on a diamond lattice.

One of the authors (M.P) acknowledges Grant from Council of Scientific and Industrial Research, India:    CSIR no.9/532(19)/2003-EMR-I. 

$^*$Author for correspondence (slnoo@magnum.barc.gov.in)


\end{document}